\documentclass[]{jpconf}
\usepackage{graphicx}
\usepackage{feynmp-auto}
\usepackage{amsmath}
\usepackage{graphicx,epsfig,float,color}
\usepackage[caption=false]{subfig}
\usepackage{hyperref}
\usepackage{calrsfs}
\usepackage{blindtext}
\usepackage[square,sort,comma,numbers]{natbib}
\usepackage{subfig}

\newcommand{\pT}{\ensuremath{p_{\text{T}}}\xspace}
\newcommand{\HT}{\ensuremath{H_{\text{T}}}\xspace}

\newcommand*{\HERWIGpp}{\textsc{Herwig}++\xspace}

\newcommand*{\POWHEG}{\textsc{Powheg}\xspace}
\newcommand*{\POWHEGBOX}{\textsc{Powheg-Box}\xspace}
\newcommand*{\POWPYTHIA}{\POWHEG{}+\PYTHIA}

\newcommand*{\PYTHIA}{\textsc{Pythia}\xspace}

\newcommand*{\SHERPA}{\textsc{Sherpa}\xspace}

\newcommand*{\RIVET}{\textsc{Rivet}\xspace}

\newcommand*{\MGPY}{MG5\_aMC@NLO+\PYTHIA8\xspace}

\newcommand*{\twothree}{\ensuremath{2\to3}\xspace}
\newcommand*{\twofour}{\ensuremath{2\to4}\xspace}

\begin{document}
\title{Performance of various event generators in describing multijet final states at the LHC}

\author{Stefan von Buddenbrock}
\address{School of Physics, University of the Witwatersrand, Johannesburg 2050, South Africa.}
\ead{stef.von.b@cern.ch}

\begin{abstract}
At the Large Hadron Collider (LHC), the most abundant processes which take place in proton-proton collisions are the generation of multijet events.
These final states rely heavily on phenomenological models and perturbative corrections which are not fully understood, and yet for many physics searches at the LHC, multijet processes are an important background to deal with.
It is therefore imperative that the modelling of multijet processes is better understood and improved.
For this reason, a study has been done with several state-of-the-art Monte Carlo event generators, and their predictions are tested against ATLAS data using the \RIVET framework.
The results display a mix of agreement and disagreement between the predictions and data, depending on which variables are studied.
Several points for improvement on the modelling of multijet processes are stated and discussed.
\end{abstract}

\section*{Introduction}

The biggest challenges to deal with in proton-proton ($pp$) collisions arise from multijet processes, as far as Standard Model (SM) backgrounds are considered.
Due to the nature of quantum chromodynamics (QCD), multijet production processes have the largest cross sections at the Large Hadron Collider (LHC).
In addition to this, their partial reliance on non-perturbative QCD makes them difficult to deal with from a theoretical perspective.
This is because simulation of fragmentation and hadronisation depend on a non-perturbative calculations, these often being done using phenomenological models.
It is therefore of importance to study the performance of event generators in describing multijet final states, since certain combinations of matrix element (ME) calculations, parton shower (PS) and hadronisation models do not always provide an accurate description of the data.

In ATLAS, a number of generators are used to model multijet processes.
These are discussed in detail in the next section.
The predictions of these generators can be compared both to each other and to data corrected for detector effects (unfolded datasets).
The simplest way of doing this is by using the \RIVET analysis system~\cite{Buckley:2010ar}, which has a large set of built in analyses and distributions of unfolded data from various experiments.
This short paper will present a subset of distributions relating to multijet processes, and compare the current set of ATLAS Monte Carlo (MC) multijet samples to unfolded data.
From these results, information can be extracted about how to improve the modelling of the generators for future generation of samples in ATLAS.

\section*{Multijet event generators in ATLAS}

A variety of MC event generators are used for studying multijet topologies in ATLAS.
These involve different combinations of ME and PS programs.
For a general review of event generators currently used in LHC physics, the reader is encouraged to look at Ref.~\cite{Buckley:2011ms}.
Below is a list of the event generators considered in this study, as well as a few notes about their set up:
\begin{itemize}
\item \PYTHIA 8~\cite{Sjostrand:2014zea}: The prediction by \PYTHIA 8 is sliced up by jet \pT using filters. The lowest \pT filtered samples use the \PYTHIA 8 built in diffractive scattering processes (\texttt{SoftQCD}) to generate events. The rest of the slices use the elastic scattering processes (\texttt{HardQCD}). The chosen tune for the \PYTHIA 8 samples is the A14 tune~\cite{ATL-PHYS-PUB-2014-021}, which assumes the \texttt{NNPDF23LO} parton density function (PDF).
\item \SHERPA~\cite{Hoeche:2012yf}: The official \SHERPA samples make use of a \twothree ME calculation,\footnote{That is, up to three partons can be generated in the final state.} matched with a CKKW scheme to a default \SHERPA PS that use the CT10 tune. This sample has known issues with forward jets. The \SHERPA prediction is also sliced in jet \pT.
\item \POWPYTHIA 8~\cite{Alioli:2010xd}: The \POWHEG ME is generated using the \texttt{Dijet} code that is provided with version 2 of the \POWHEGBOX. It is passed to the \PYTHIA 8 PS, with the A14 tune. The sample is also sliced in jet \pT.
\item \HERWIGpp~\cite{Bahr:2008pv}: Like \PYTHIA 8, the \HERWIGpp sample makes use of the built-in \texttt{MEMinBias} process to simulate diffractive scattering for the lowest two slices in jet \pT, and the \texttt{MEQCD2to2} for the remaining slices. These samples make use of the UE-EE5 tune, and therefore the \texttt{CTEQ6L1} PDF.
\item \MGPY~\cite{Alwall:2014hca}: The \MGPY samples use a \twofour ME matched with a \PYTHIA 8 PS using the CKKW-L scheme. The ME makes use of the \texttt{NNPDF30NLO} PDF, while the PS uses the A14 tune as described above. These samples are sliced at the ME level in parton \HT.
\end{itemize}

\section*{Key comparisons to data}

As mentioned above, \RIVET is used to compare the predictions of these variables against each other and unfolded data.
In this short paper, the predictions are compared in the context of three different aspects of jet physics, namely azimuthal decorrelations, jet fragmentation and jet shapes.
Note that all jets considered in the following analyses are constructed using the anti-$k_\text{T}$ algorithm~\cite{Cacciari:2008gp} with a radius parameter of $R=0.6$.

\subsection*{Azimuthal decorrelations}

Purely elastic scattering of QCD partons most often results in a dijet event -- that is, exactly two well separated jets in the final state.
In such a case, the azimuthal separation between the two jets should be $\pi$ radians.
However, in theory, one expects to see more QCD interactions in the elastic scattering of quarks and gluons.
This extra activity can produce more jet activity in multijet events.
Depending on how much more activity is found in the event, the azimuthal angle between the two leading jets will deviate from $\pi$.
This is known as an azimuthal decorrelation.

\begin{figure}[t]
  \includegraphics[width=0.49\textwidth]{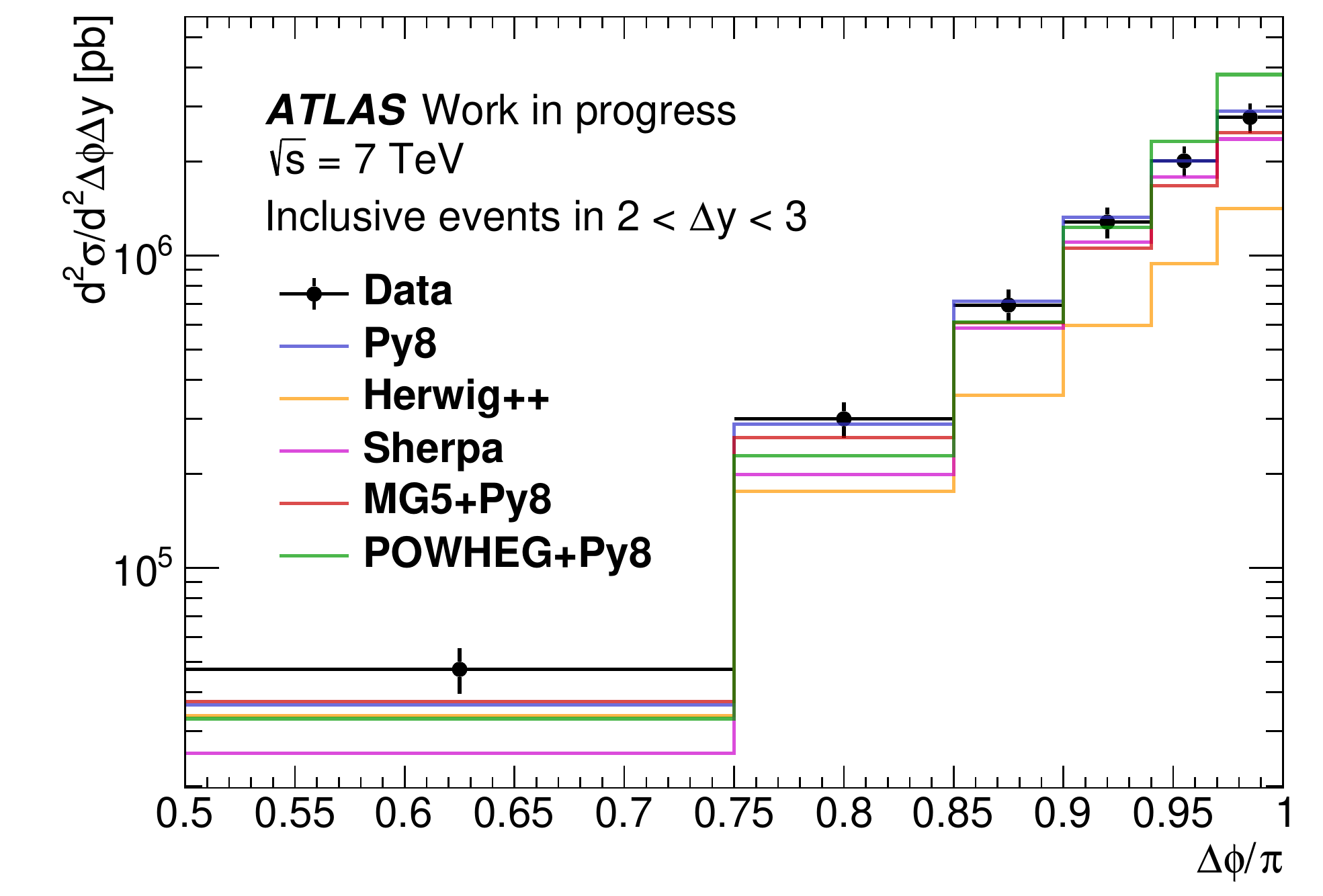}
  \includegraphics[width=0.49\textwidth]{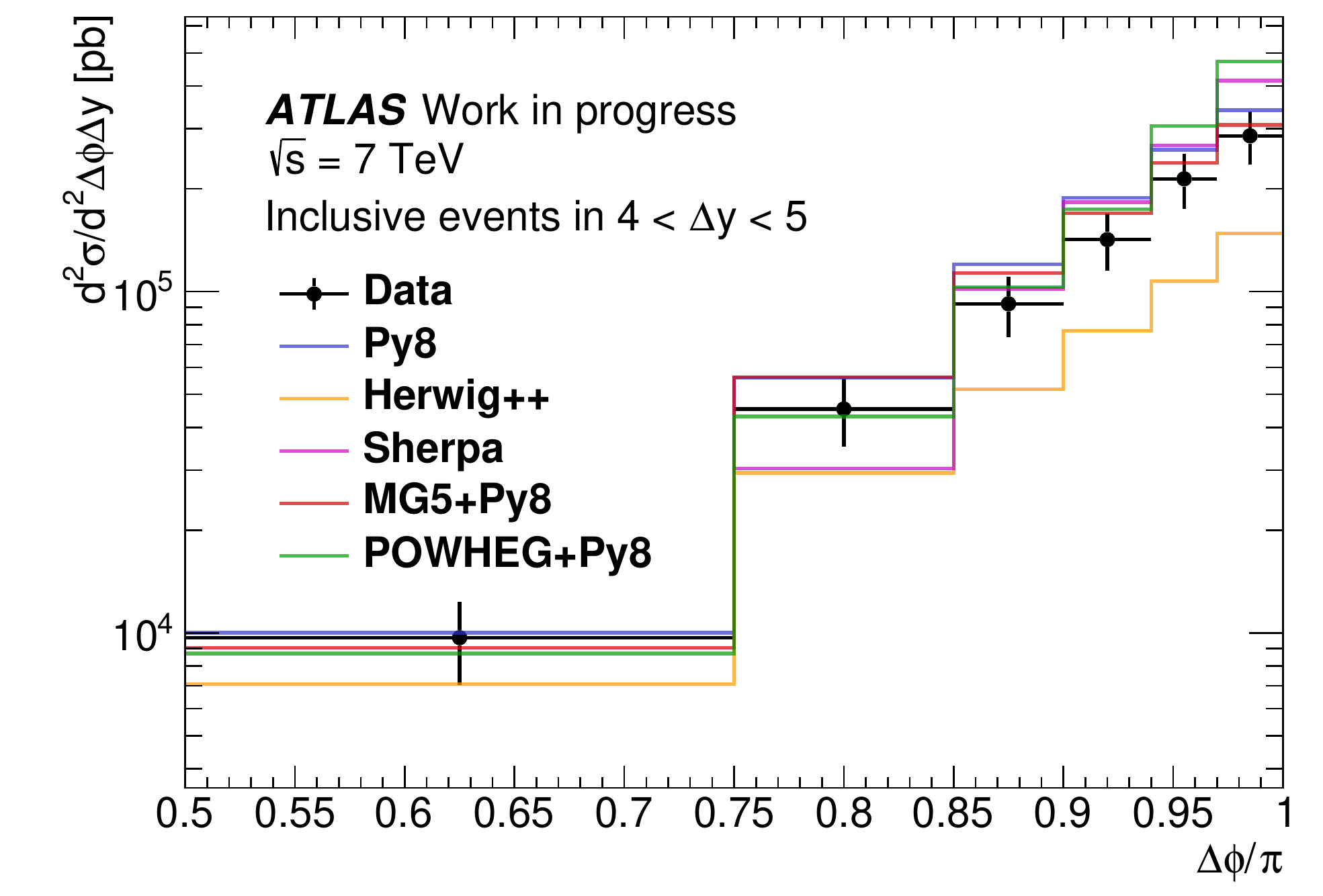}
  \caption{Differential cross section measurements as a function of the azimuthal angle between the two leading jets in multijet events~\cite{STDM-2012-17}. The plots are made for inclusive multijet events in bins of rapidity separation, with $2<\Delta y<3$ on the left and $4<\Delta y<5$ on the right.}
  \label{fig:AD}
\end{figure}

In order to study azimuthal decorrelations, one typically looks at the azimuthal angle between the two leading jets in multijet events.
ATLAS performed a differential cross section measurement of azimuthal decorrelation variables with the Run 1 7~TeV dataset~\cite{STDM-2012-17}.
The corresponding \RIVET routine for this analysis is \texttt{ATLAS\_2014\_I1307243}.
In \autoref{fig:AD}, some comparison plots are shown from this analysis.
The different generators mostly perform well against the data, although discrepancies arise in different regions of the distributions, particularly for \HERWIGpp.

\subsection*{Jet fragmentation}

The behaviour of the fragmentation function used in different PS models is most commonly studied by looking at the densities of jet constituents in selected jets.
The simplest \RIVET routine to use when studying jet fragmentation is \texttt{ATLAS\_2011\_I929691}, which is a 7~TeV measurement of charged jet constituent densities as a function of three different variables~\cite{STDM-2011-14}.
Firstly, the variable $z$ is scanned, which is the fraction of longitudinal momentum carried by a jet constituent:
\begin{equation}
z=\frac{\vec{p}_\text{jet}\cdot\vec{p}_\text{ch}}{\left|\vec{p}_\text{jet}\right|^2}.
\end{equation}
Here, $\vec{p}_\text{ch}$ denotes the 3-momentum of the charged jet constituent, and $\vec{p}_\text{jet}$ is the 3-momentum of the jet.
Secondly, the distance between the jet axis and the jet constituent in units of $\phi$ and $y$ is scanned over (denoted by $r$).
And thirdly, the jet constituent's momentum transverse to the jet axis, $p_\text{T}^\text{rel}$, is scanned over:
\begin{equation}
p_\text{T}^\text{rel}=\frac{\left|\vec{p}_\text{ch}\times\vec{p}_\text{jet}\right|}{\left|\vec{p}_\text{jet}\right|}.
\end{equation}

\begin{figure}[t]
  \includegraphics[width=0.49\textwidth]{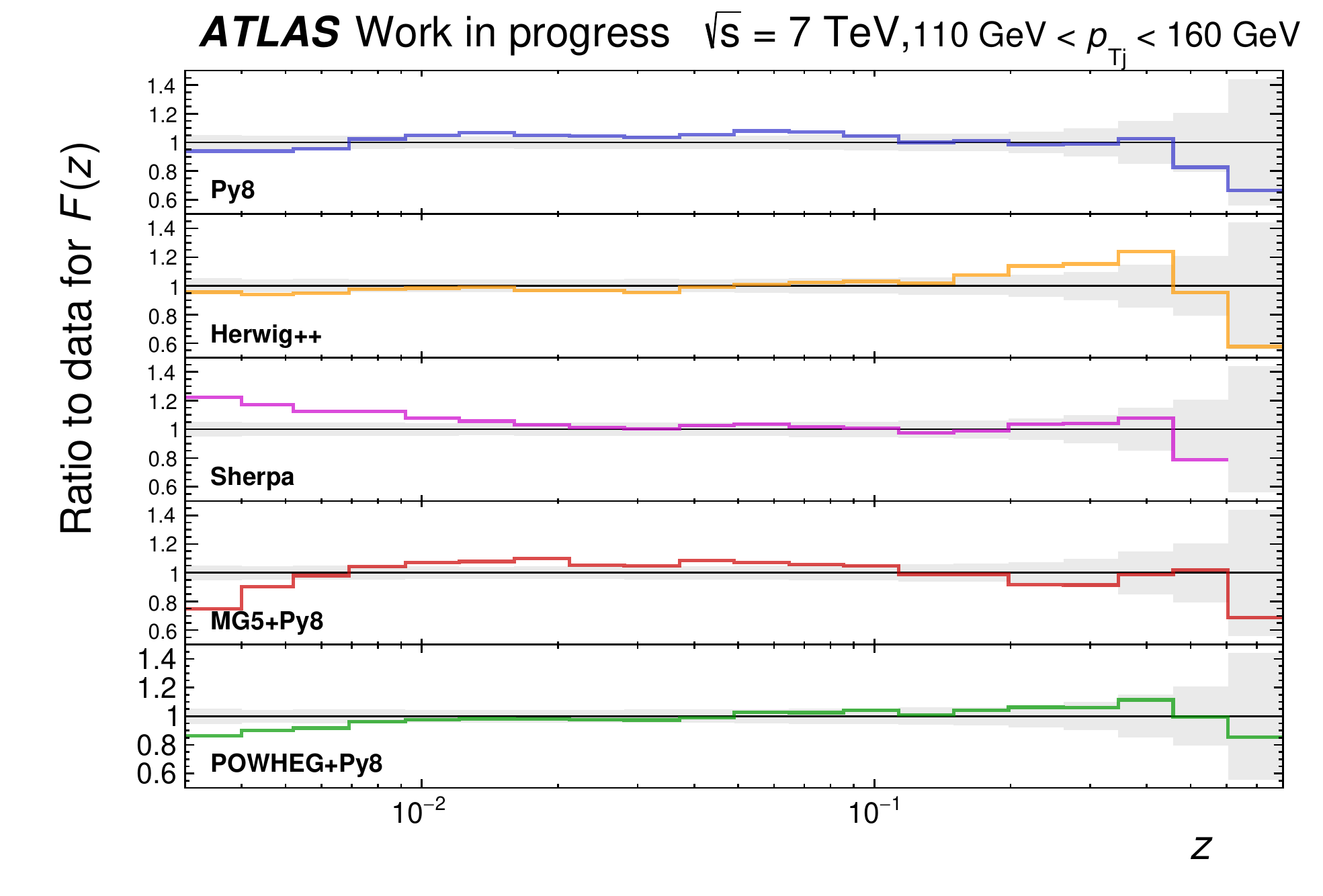}
  \includegraphics[width=0.49\textwidth]{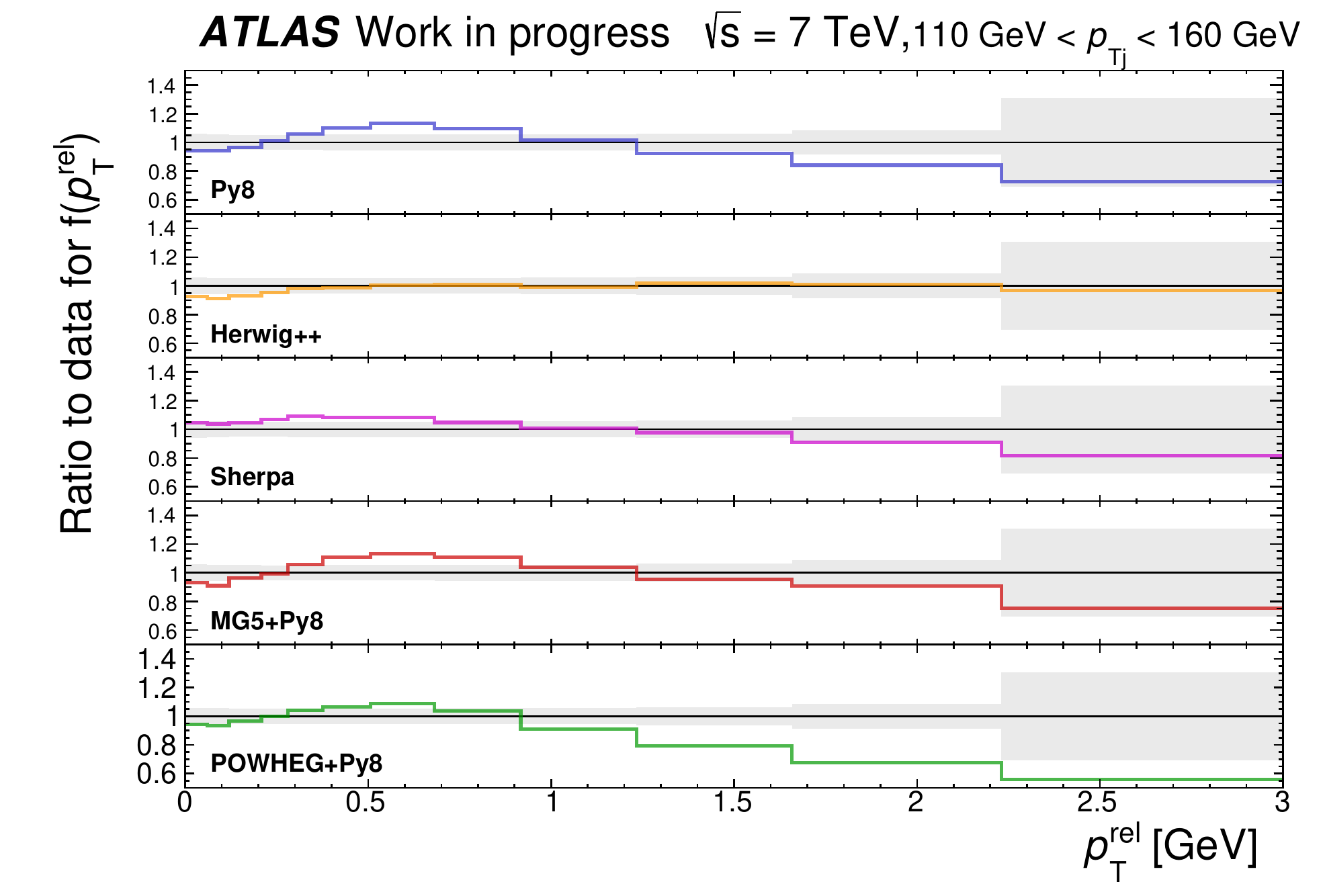}
  \caption{Ratio plots of the different event generators compared to the data in measurements of jet constituent densities as a function of $z$ (left) and $p_\text{T}^\text{rel}$ (right)~\cite{STDM-2011-14}. These plots are both shown in the same bin of jet \pT.}
  \label{fig:JF}
\end{figure}

In \autoref{fig:JF}, plots are shown for jet constituent densities as a function of two of these variables.
In this case, \HERWIGpp arguably performs the best compared with the data, and \SHERPA tends to perform the poorest.
This is most probably due to the old version of \SHERPA used by ATLAS in the official samples.
Recent studies on newer \SHERPA samples in ATLAS have seen the problems with jet fragmentation fixed, although these results could not be shown in this short paper.

\subsection*{Jet shapes}

Jet algorithms can tell us about the geometry of the constituents of a jet.
But to understand how energy is distributed in the average jet, it is more instructive to look at jet shapes.
Similarly to jet fragmentation measurements, jet shapes are studied through looking at jet constituent densities.
These are distributed as a function of the distance away from the axis of a jet, $r$.
Typically, we look at the jet \pT weighted density in bins of annulus areas in the jet cone,
\begin{equation}
  \rho(r)=\frac{1}{\Delta rN_\text{jet}}\sum_\text{jets}\frac{\pT(r-\Delta r/2, r+\Delta r/2)}{\pT(0,R)},
\end{equation}
where $p_\text{T}(r_1,r_2)$ is the sum of the jet constituent \pT between $r_1$ and $r_2$ away from the jet cone axis.
In addition to this, we measure the integrated \pT weighted density,
\begin{equation}
  \Psi(r)=\frac{1}{N_\text{jet}}\sum_\text{jets}\frac{\pT(0,r)}{\pT(0,R)}.
\end{equation}

\begin{figure}[b]
  \includegraphics[width=0.49\textwidth]{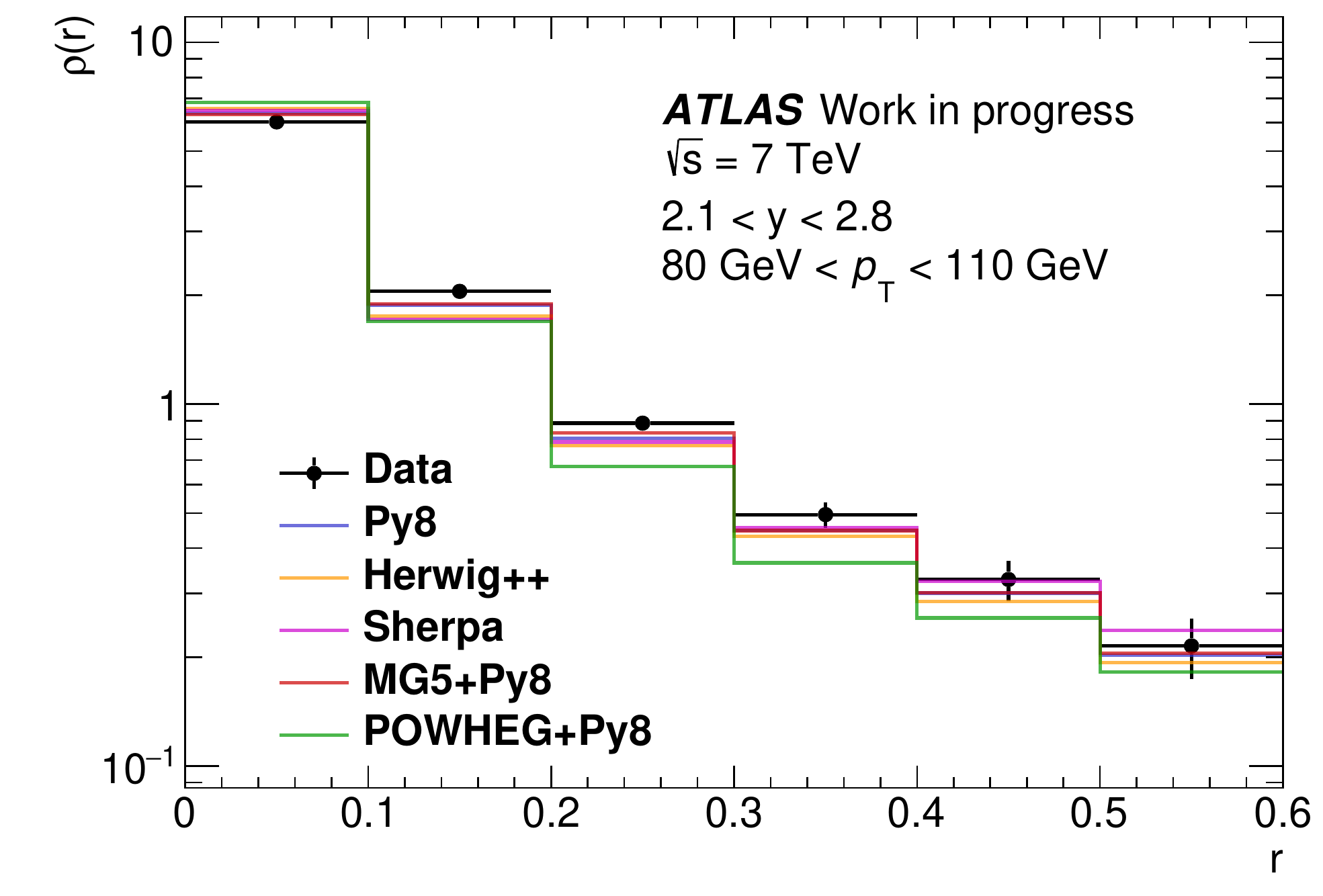}
  \includegraphics[width=0.49\textwidth]{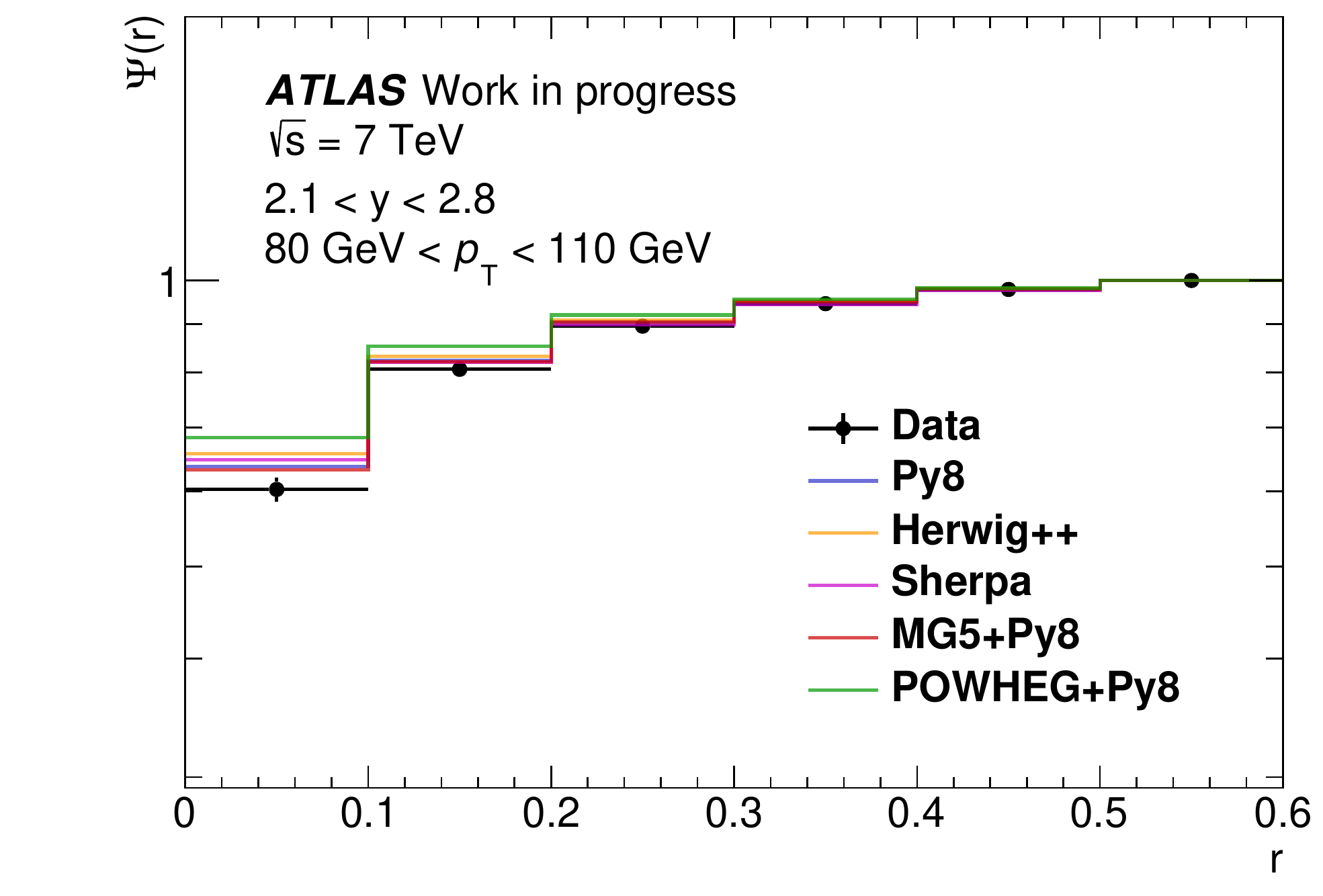}
  \caption{Jet shape measurements in terms of differential \pT density (left) and integrated \pT density (right)~\cite{STDM-2010-10}.}
  \label{fig:JS}
\end{figure}

The \RIVET routine \texttt{ATLAS\_2011\_S8924791} contains a large set of doubly differential jet shapes corresponding to an ATLAS 7~TeV measurement~\cite{STDM-2010-10}.
In \autoref{fig:JS}, some plots are shown in a single bin of the ATLAS analysis.
Here, most of the generators considered agree relatively well with the data.
It should be noted that \POWPYTHIA seems to predict a different jet shape than what is seen in the data.

\section*{Summary}

Using the ATLAS multijet samples, comparisons have been made to unfolded data using \RIVET.
Three different measurements have been considered in this short paper.
In each, it can be seen that the different generators tend to perform better in some regions of the phase space than others, while there is no clear choice for one generator performing systematically better than any of the others.

However, the information from these comparisons is still useful for the ATLAS collaboration to improve the modelling of multijet processes by knowing where the current predictions fail.
There are yet many more measurements that can be considered in this study, and in future these studies will be extended to a more comprehensive study.

%\section*{References}

\bibliographystyle{iopart-num}
\bibliography{ref.bib}

\end{document}